\documentstyle[epsf,twocolumn,prl,aps]{revtex}
\begin{document}

\title{Rate theory for correlated processes: Double--jumps in adatom
  diffusion.}

\author{Joachim Jacobsen$^1$, Karsten W. Jacobsen$^{1,2}$, and 
James P. Sethna$^1$}

\address{${}^1$Laboratory of Atomic and Solid State Physics, Cornell
  University, Ithaca, NY 14853-2501, USA\\
${}^2$Center for Atomic-scale Materials Physics, Department of Physics,\\
Technical University of Denmark, DK-2800 Lyngby, Denmark}

\date{\today}

\maketitle

\begin{abstract}
We study the rate of activated motion over multiple barriers, in particular
the correlated double--jump of an adatom diffusing on a missing-row
reconstructed Platinum (110) surface.  We develop a Transition Path Theory,
showing that the activation energy is given by the
minimum--energy trajectory which succeeds in the double--jump.
We explicitly calculate this trajectory within an effective-medium molecular
dynamics simulation.  A cusp in the acceptance region 
leads to a $\sqrt{T}$ prefactor for the activated rate of double--jumps.
Theory and numerical results agree.

\end{abstract}


Reaction rates and diffusion rates in crystalline environments
typically have an Arrhenius dependence on temperature $r_1 = \nu
\exp{(-E_{\rm TS}/T)}$.  This asymptotic rate at low temperatures may
be calculated using Transition State Theory (TST)\cite{tst} , where
$E_{\rm TS}$ is the energy of the saddle-point atomic configuration
separating the initial and final states (the Transition State), and
$\nu$ is a temperature--independent prefactor involving the curvatures
of the energy surface.

Many rates and transitions are not described by a simple crossing of a
single barrier, and we should expect that their rates will not be
given by the simple Arrhenius form.  In this paper we study a
double--jump: a correlated diffusion event where an atom crosses two
barriers.  The development of field--ion microscopy (FIM) and scanning
tunneling microscopy (STM) has made it possible to track the motion of
individual atoms at surfaces~\cite{fim-stm}, and to directly measure
the rates of these correlated transitions~\cite{ehrlich,aarhus}.
Developing what we call Transition Path Theory (TPT), we show the rate
is determined by the energy $E_{\rm TP}$ of the transition path
(minimizing the energy among all paths which succeed in the
double--jump), and has the asymptotic form $r_2 = C \sqrt{T}
\exp{(-E_{\rm TP}/T)}$.  We develop an efficient numerical method to
calculate this minimum-energy path and the rate, and use it to
describe double--jumps along the troughs of the missing-row
reconstructed Pt(110) surface within effective--medium
theory~\cite{emt} --- making contact with the Arrhenius double--jump
rate measured in the recent {\AA}rhus experiment\cite{aarhus}.

Previously, the rates of double--jumps have been discussed
theoretically as so-called dynamical corrections to transition state
theory\cite{voter,metiu} or in the context of generalized Langevin
equation models\cite{pollak,ferrando}.

\begin{figure}[thb]
  \begin{center}
    \leavevmode
    \epsfxsize=7cm
    \epsffile{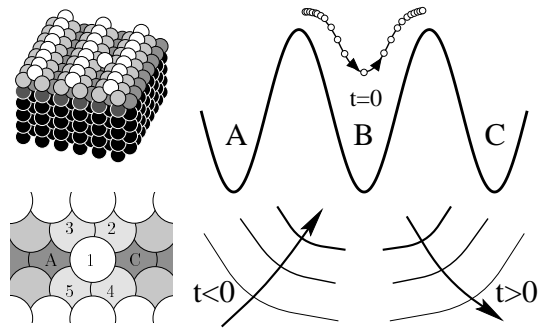}
  \end{center}

\caption{The geometry of the Pt(110)-surface with an adatom
  and the Transition Path (TP) for a double--jump.  At upper left, one
  sees the actual volume of atoms used in the simulations, with
  periodic boundary conditions and static atoms at the bottom.  Lower
  left shows the geometry of the system with the adatom \#1 in a
  valley formed by the reconstruction.  At right we show schematically
  the TP and the energy.  The TP starts at $t=-\infty$ with the system
  locally (close to the adatom) at the transition state (TS) potential
  energy saddle-point between wells A and B; in addition, there is an
  energy $\Delta E = E_{\rm TP}-E_{\rm TS}$ stored in the degrees of
  freedom far away. The additional energy radiates in from infinity
  and helps the adatom cross the well and get to the second TS.  Our
  calculation shows that the adatom rolls off the hill, hits atoms
  2\&4 and knocks them aside, then is hit from behind by atoms 3\&5,
  who boost the adatom up the other hill. As $t \to \infty$ the energy
  $\Delta E$ is again radiated away to infinity
  \protect\cite{LocalMode}.}
\label{fig:geom}
\end{figure}

Consider an extra adatom on the reconstructed (110) surface
(Fig.~\ref{fig:geom}).  A single--jump diffusion event is a thermal
fluctuation in which the adatom moves along the troughs in the surface
from one potential well over the transition barrier to a neighboring
well (say from well A to B in the right-hand Fig.~\ref{fig:geom}) .
In a double--jump, the atom moves from one potential well over two
barriers before settling down in a well (A to C in
Fig.~\ref{fig:geom}).  The double--jumps can be distinguished from two
subsequent single--jumps only if there is a separation of time scales:
the duration time for a double--jump to pass through the central well
(roughly a picosecond for Pt) has to be much smaller than the inverse
rate $1/r_1$ for single jumps (1.4 nanoseconds at the highest
temperatures we work at).

In the case of a single--jump the transition state energy $E_{\rm TS}$
is the smallest possible thermal fluctuation for a single--jump to
occur.  For our calculation, $E_{\rm TS}$ is $469.5$ meV.  What is the
smallest possible thermal fluctuation, in which the adatom performs a
double--jump: fluctuating to the top of one hill, sliding down, up to
the top of a second hill, and then down?  Here the energy of the
minimal fluctuation involves both kinetic and potential energy: we
must look in phase space for the {\it transition path} (TP): the
lowest energy trajectory which will give us a double jump.

Before we more formally explain how the TP is calculated we shall
describe the qualitative features of the path. Consider first the time
evolution of the system if it is started out in a TS and the adatom is
given a slight push. The adatom will then slowly leave the saddlepoint
and slide down the hill into a well but it will not make it over the
next barrier because energy will be transferred to other atoms in the
system.  Eventually the adatom will be at rest at the bottom of the
well and all the energy $E_{\rm TS}$ will be in the degrees of freedom
far away from the adatom. The TP is defined so that it brings the
adatom to the top of the next barrier (the next TS) with the minimum
additional energy ($\Delta E = E_{\rm TP}-E_{\rm TS}$) necessary. In
the TP the system is at $t=-\infty$ at the TS for all the degrees of
freedom close to the adatom, but an energy $\Delta E$ is stored in the
degrees of freedom far away. As the adatom slides down the hill the
energy is radiating in from infinity helping the adatom up to the
second TS (Fig.~\ref{fig:geom}). As $t \to \infty$ the adatom (and all
local coordinates/momenta) approaches the second TS and the energy
$\Delta E$ is again radiated away to infinity \cite{LocalMode}.  The
additional energy $\Delta E$ results in a higher activation energy for
double--jumps than for single--jumps.

In finding the minimum energy double--jump path, it is important to
use as variables the positions and momenta $\Gamma$ at $t=0$ when the
adatom is crossing the bottom of the central well, because at this
time the additional energy $\Delta E$ is most localized near the
adatom.  Let $p_x$ be the $x$-momentum of the adatom at $t=0$, where
$x$ is the coordinate along the trough.  To find $\Gamma_{\rm TP}$ of
the TP, we specify the positions and the momenta of the nearby atoms
within a radius $R$ of the adatom, and place atoms outside $R$ at
their relaxed positions.  We then vary $p_x$ to find $p_x^{min}$,
where the adatom just barely succeeds in the double--jump: crosses
into the well on the right at large times, and into the well on the
left at large negative times\cite{pxmin}. We calculate the total
energy, and minimize with respect to the positions and momenta of the
nearby atoms.  With all atoms in their relaxed positions, 652.0 meV of
energy is needed in $p_x$ to get a double--jump.  Optimizing degrees
of freedom within $R=$ 1, 2.2, 3.1, 3.8, and 4.9 nearest neighbor
distances of the adatom, the total energy decreases to 621.7, 595.8,
591.4, 590.2 and 589.8 meV respectively.  The energy converges rapidly
as more degrees of freedom are optimized and $\Gamma \to \Gamma_{\rm
  TP}$.  We estimate $E_{\rm TP}$ to be 589.8 meV: 120.3 meV higher
than the single--jump $E_{\rm TS}$.  The calculated TP appears to have
both of the obvious possible symmetries: one reflection plane, and one
reflection plane plus time-reversal.  The time-reversal symmetry
implies that the local atoms both start and end at zero velocity at
the saddle points.

\begin{figure}[htb]
  \begin{center}
    \leavevmode
    \epsfxsize=7cm
    \epsffile{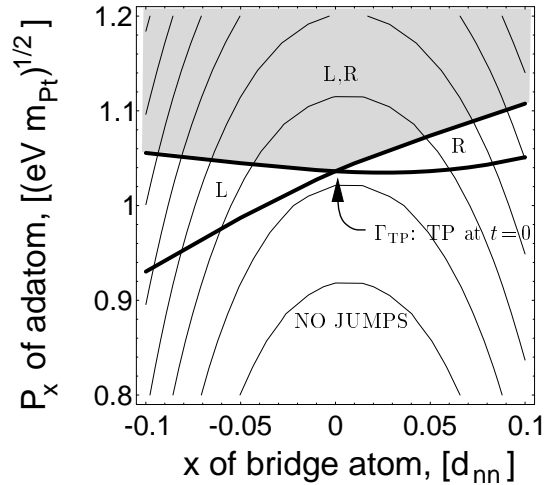}
  \end{center}
    \caption{Energy contours (100 meV) in a cut in phase space. 
      $\rm m_{\lower1pt\hbox{$\scriptstyle \rm P\! t$}}$ and $\rm
      d_{\rm n\! n}$ are the atomic mass and nearest neighbor distance
      of Platinum.  The bridge atom is atom 2 in Fig.~1.  All other
      coordinates are fixed at their values in the Transition Path
      (TP) at time $t=0$. L(R) indicates regions in phase space, where
      the adatom makes a jump to the left(right), and the solid lines
      are trajectories where the local atoms end at the saddlepoints.
      The double--jump region (gray shaded) is above both lines, and
      so the bounding surface for the double--jump region has a cusp
      passing through $\Gamma_{\rm TP}$, the phase-space point taken
      by the TP at $t=0$.  }
    \label{fig:phasespace}
\end{figure}

To calculate the rate at low temperatures, we need to sum over all the
low-energy double--jump trajectories.  Fig.~\ref{fig:phasespace} gives
an idea about how the double--jump region in phase space looks.  The
figure shows this region in a 2-dimensional cut in phase space (the
plane spanned by the $x$-coordinate of a bridge atom (atom 2 in
Fig.~\ref{fig:geom}) and $p_x$ of the adatom; all the remaining
coordinates are at their $\Gamma_{\rm TP}$ values).  {\bf R} is the
region corresponding to trajectories where the adatom will jump over
the barrier to the right; {\bf L} is the region where the adatom came
over the barrier from the left.  The TP is the lowest-energy
trajectory which came from the left and makes it to the right, and is
situated on the cusp at the (co-dimension two) intersection of the
bounding surfaces of {\bf R} and {\bf L}.

Our method for calculating the rate of double--jumps is analogous to
the standard TST method for calculating the rates for single--jumps.
TST expresses the hopping rate as the flux through the dividing
surface; at low temperatures it is well described as a harmonic
expansion of the energy about the TS~\cite{tst}:
\begin{eqnarray}
  \label{eq:r1}
  r_1^{\rm TST} & = & \langle \Theta(v_r) v_r
       \delta(x_r-{x_r}^b)\rangle/ Z_W \nonumber \\
       & = & \frac{1}{2 \pi} 
   \frac{\prod_{i=1}^{3N} \omega_i}{\prod_{i=1}^{3N-1} 
        \omega_i^{\rm TS}} \exp (-E_{\rm TS}/T).
\end{eqnarray}
In this expression $\langle \ldots \rangle$ denotes a thermal average,
$Z_W$ denotes the partition function in one well.  The {\it reaction
  coordinate} $x_r$ equals $x_r^b$ at the top of the barrier; the
$\omega$'s denote the eigenfrequencies at the bottom of the well, and
the $\omega^{\rm TS}$'s denote the eigenfrequencies found by the
harmonic expansion at the TS (excluding the imaginary, unstable
frequency of the saddle point).  Choosing the coordinate $x_r$ to be
along the eigendirection of the imaginary frequency minimizes the
recrossing corrections\cite{tst}.

Similarly, the double--jump rate can be calculated as the thermal
average of the flux through a surface $\tilde x=0$ due to double--jump
trajectories:
\begin{equation}
  \label{eq:r2}
  r_2 = \langle \Theta_{dj} v_{\tilde x} \delta(\tilde x)\rangle/ Z_W.
\end{equation}
Here $\Theta_{dj}$ is one if the trajectory is a double--jump, and
zero otherwise.  In both cases the rate is calculated as a flux
through a surface, with a $\Theta$-function keeping only trajectories
which succeed in the transition.

We generate an importance-sampling distribution of $M$ trajectories,
$\{\Gamma_\alpha\}$, in the neighborhood of the transition path by
adding $\Gamma_{\rm TP}$ to a thermal ensemble restricted to $\tilde
x=0$ at the bottom of the well. We found it important numerically to
choose $\tilde x$ to be a relative coordinate: the difference between
the $x$-coordinate of the adatom and the average $x$-coordinate of the
four bridge atoms (\#2-5 on Fig.~\ref{fig:geom}).  We then calculate
the ratio of the double--jump to single--jump rates:
\begin{eqnarray}
  \label{eq:r2overr1}
  {r_2\over r_1} & = & {\langle \Theta_{dj} v_{\tilde x} 
                \delta(\tilde x)\rangle
                \over \langle \Theta(v_r) v_r \delta(x_r\! -\! {x_r}^b)\rangle}
          =  {  \null ~~~~~~~ \langle \delta(\tilde x) \rangle ~~~
            \langle \Theta_{dj}
                v_{\tilde x} \delta(\tilde x)\rangle
                \over \langle \Theta(v_r) v_r \rangle \langle
                \delta(x_r\! -\! {x_r}^b)\rangle \langle \delta(\tilde x)
                \rangle} \nonumber \\ 
          & = & \sqrt{2 \pi m / T} \left[ e^{E_{\rm TS}/T} 
                \prod_{i=1}^{3N-1} {\omega_i^{\rm TS} \over
                  \tilde\omega_i} 
                  \right] \\
              & & \times  {1\over M} \sum_{\alpha=1}^{M} v_{\tilde x,\alpha}
                \Theta_{dj}(\Gamma_\alpha)
                e^{(E(\Gamma_\alpha-\Gamma_{\rm TP})-
                E(\Gamma_\alpha)) / T}. \nonumber  
\end{eqnarray}
We determine which trajectories are double--jumps numerically: we run
the molecular dynamics trajectory forward and backward in time for a
few picoseconds until the trajectory either recrosses the original
well (failure, $\Theta_{dj}=0$) or crosses the bottom of both of the
adjacent wells (success, $\Theta_{dj}=1$).  This expression is
complicated by the fact that we are measuring the flux for
single--jumps at the top of the barrier, and the flux for
double--jumps at the bottom of the well: the term in square brackets
is precisely the ratio of probabilities of being at these two planes,
in the harmonic approximation (the $\tilde \omega$'s being the
eigenfrequencies in the plane $\tilde x=0$, and we evaluate $\prod
{\omega_i^{\rm TS} / \tilde\omega_i}$ to be 1.722).  Our double--jump
term is fully nonlinear.

\begin{figure}[htb]
  \begin{center}
    \leavevmode
    \epsfxsize=7cm
    \epsffile{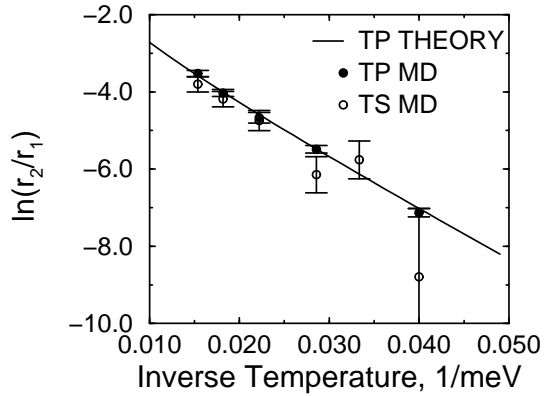}
  \end{center}
    \caption{Arrhenius plot of the rate of double--jumps relative to single
      jumps, showing the Transition State Molecular Dynamics (TS MD),
      Transition Path Molecular Dynamics (TP MD), and Transition Path
      Theory results. The TP Theory curve is $r_2/r_1 = C_2 \sqrt{T}
      \exp{-(E_{\rm TP}-E_{\rm TS})/kT}$, where the one parameter
      $C_2$ is fitted to the TP MD data to be 0.022 meV$^{-1/2}$.}
    \label{fig:hightemp}
\end{figure}

Fig.~\ref{fig:hightemp} shows the ``TP MD'' values calculated using
equation~\ref{eq:r2overr1}.  Also shown are rate calculations
performed using the more traditional method described by Voter and
Doll\cite{voter} (transition state molecular dynamics ``TS MD''),
where trajectories are started with a thermal distribution at the TST
dividing surface, and the fraction of double--jumps is directly
measured.  Within the statistical error bars the two methods give the
same rate, as they should.  However, the uncertainty in the
traditional rate determination increases drastically at lower
temperatures because the fraction of double--jump trajectories in the
thermal ensemble becomes small.  In our language, the old method
centers attention not at $\Gamma_{\rm TP}$, but far away at the
single--jump transition state.

We now discuss the asymptotic behavior of the double--jump rates at
low temperatures.  In analogy to the harmonic expansion around the TS
which lead to the TST rate with an Arrhenius behavior for the
single--jump rate (Eq.~\ref{eq:r1}) we can define a transition path
theory rate $r_2^{\rm TPT}$ for the double--jumps by performing an
expansion of equation~\ref{eq:r2} around the TP.

$\Theta_{dj}$ is a Heaviside step function in
$p_x-p_x^{min}(\tilde\Gamma)$, (Fig.~\ref{fig:phasespace}), where
$\tilde\Gamma$ is the phase space point at $t=0$ excluding $x$ and
$p_x$ of the adatom.  The $x$ and $p_x$ parts of the phase-space
integral in the numerator of Eq.~\ref{eq:r2} (choosing here for
convenience $\tilde x = x$) can be explicitly carried out, leading to
\begin{equation}
  \label{eq:r2int}
  r_2 = \frac{T}{Z_W} \int d\tilde\Gamma\,
  \exp(-\tilde E(\tilde\Gamma)/T).
\end{equation}
where $\tilde
E(\tilde\Gamma)=E(x=0,p_x^{min}(\tilde\Gamma),\tilde\Gamma)$ is the
energy on the boundary of the double--jump region.  In the
low-temperature limit a harmonic expansion of $\tilde E(\tilde\Gamma)$
around the TP in the $6N\!-\!2$ variables $\tilde\Gamma$ can be
performed and the resulting multidimensional Gaussian integral can in
principle be carried out.  Focusing on the temperature dependence of
the prefactor, we note that $p_x^{min}$, and hence $\tilde E$ has a
cusp in one degree of freedom (see Fig.~\ref{fig:phasespace}), and
that there is a $6N-3$--dimensional subspace, where the energy is
presumably locally quadratic. These $6N-3$ degrees of freedom should
give rise to a factor $\sqrt{T}$ (each) in the prefactor, whereas the
cusp degree of freedom gives by integration a factor $T$, so overall
we get the prefactor $T\times T^{(6N-3)/2} \times T / T^{3N} =
\sqrt{T}$. The TPT rate for the double--jumps is therefore of the form
\begin{equation}
  \label{eq:tpt}
  r_2^{\rm TPT} = C_1 \sqrt{T} \exp(-E_{\rm TP}/T),
\end{equation}
with $C_1$ a temperature--independent constant. Alternatively we may
write $r_2^{\rm TPT}/r_1^{\rm TST} = C_2 \sqrt{T} \exp(-(E_{\rm
  TP}-E_{\rm TS})/T)$, where $C_2$ is the constant $C_1/\nu$.

The TPT-rate expression is shown as the solid curve in
Fig.~\ref{fig:hightemp} with only the constant $C_2$ fitted to the
calculated rates. The TPT rate is clearly in agreement with the
simulation results even up to the highest temperature $T= 65$ meV.  At
high temperatures corrections caused by anharmonic effects can be
expected and higher order multiple jumps (triple, quadruple, etc.) may
also play a role in the diffusivity.  It is easy to fit the TP MD data
of Fig.~\ref{fig:hightemp} with an Arrhenius form with constant
prefactor, but it yields the wrong energy barrier (139 meV, 16\% too
high).

\begin{figure}[htb]
  \begin{center}
    \leavevmode
    \epsfxsize=7cm
    \epsffile{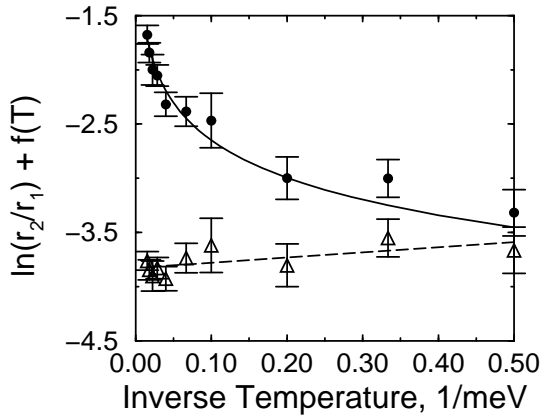}
  \end{center}
    \caption{Temperature dependence of TP MD data for the rate
      of double--jumps relative to single--jumps. The black dots show
      the data corrected for the predicted exponential temperature
      dependence ($f(T) = \Delta E/T$, $\Delta E = E_{\rm TP}-E_{\rm
        TS}$ = 120.3 meV). The curvature of this set of points proves
      that the prefactor depends on temperature, and is well described
      by the solid line, which is a plot of $\ln{C_2 \sqrt{T}}$, where
      $C_2$ is 0.022 meV$^{-1/2}$ as in Fig.~3.  The open triangles
      shows the data now also corrected for the $\sqrt{T}$-dependence
      ($f(T) = \Delta E/T - \ln{T}/2$). TP Theory predicts this set of
      points this to be a constant ($\ln{C_2}$), consistent with the
      data. Linear regression (dashed line) gives a slope of $+0.47
      \pm 0.2$~meV $\sim 0.4 \%$ of $\Delta E$.  This slope is a
      reasonable estimate of how much further our estimated $E_{\rm
        TP}$ could be lowered by further optimization.}
    \label{fig:sqrttemp}
\end{figure}

Since we sample the vicinity of the TP in our numerical method, we can
extend the calculation of the double--jump rate to much lower
temperatures to confirm the theoretical asymptotic behavior.  (The
double--jump rate becomes very small.) In Fig.~\ref{fig:sqrttemp} we
show the result for $r_2/r_1$ dividing out first the predicted
Arrhenius dependence (solid symbols), and further the predicted
$\sqrt{T}$--dependence (open symbols), ending up with a set of data
points consistent with a constant. Hence our numerical method confirm
the TPT rate as given by Equation~\ref{eq:tpt}.

Comparing with the recent experiments\cite{aarhus} on the
Pt/Pt(110)-system we note that in our model the barrier for single
jumps is $E_{\rm TS} = 469.5$ meV while the activation energy
experimentally is found to be around 0.8 eV. It is well known that the
EMT potential tends to underestimate diffusion barriers for
Pt\cite{joachim}.  However, the TPT analysis confirms the experimental
observation of a thermally activated form; the calculated additional
activation energy for double--jumps $E_{\rm TP} - E_{\rm TS}= 120$ meV
is in good agreement with the experimentally determined value which is
of the order 0.1 eV.

We finally note that we have applied the TPT to the traditional
Langevin equation (Ohmic damping) previously used to describe
double--jumps\cite{pollak}.  Here the TP becomes the optimal time
evolution of the external ``fluctuating'' force.  The additional
activation energy $\Delta E$ for double--jumps equals the work done by
friction in the adiabatic potential {\it only} in the low friction
limit. We observe the same $\sqrt{T}$ prefactor.

The authors   wish  to thank Flemming Besenbacher,   Trolle Linderoth,
Christopher Henley,  and Per Stoltze  for helpful conversations.  This
work was supported  by  the National  Science Foundation  through  the
Cornell Materials Science Center   NSF-DMR-9121654.  Support was  also
provided by the Department of Energy DEFG02-88-ER45364, and the Danish
National  Research Foundation  through   the Center   for Atomic-scale
Materials Physics.


\begin{references}
\bibitem{LocalMode} In our system, a small amount of energy is left in a
long-lived local mode with the adatom oscillating vertically about the TS; this
energy will also eventually decay due to anharmonicity.
\bibitem{pxmin} An efficient way to determine $p_x^{min}$ in practice is
  by maximizing the time $t$, it takes for the adatom to get to the
  dividing surface, using that $|p_x - p_x^{min}| \propto
  \exp{(-4 \pi \nu_{\lower1.5pt\hbox{$\scriptstyle \rm T\!S$}} t)}$ 
  for large $t$, where $\nu_{\lower1.5pt\hbox{$\scriptstyle \rm T\!S$}}$
  is the frequency of the unstable mode at the saddle--point.
\end{references}
\end{document}